\documentclass[letterpaper, 10 pt, conference]{ieeeconf}  

\usepackage{multirow}
\usepackage{xcolor}
\usepackage{graphicx}
\usepackage{url}
\usepackage{makecell}
\usepackage[noadjust]{cite}
\usepackage{booktabs}
\usepackage{hyperref}
\usepackage{subcaption}

\IEEEoverridecommandlockouts                              

\title{\LARGE \bf
Evaluating Zero-Shot and One-Shot Adaptation of Small Language Models in Leader-Follower Interaction
}

\author{Rafael R. Baptista$^{1}$, André de Lima Salgado$^{2}$, Ricardo V. Godoy$^{1}$, Marcelo Becker$^{1}$,\\Thiago Boaventura$^{1}$, and Gustavo J. G. Lahr$^{3}$
\thanks{$^{1}$Rafael, Ricardo, Marcelo, and Thiago are with the University of Sao Paulo, Brazil.}%
\thanks{$^{2}$André is with Federal University of Lavras, Brazil.}%
\thanks{$^{3}$Gustavo is with the Faculdade Israelita de Ensino e Pesquisa Albert Einstein, Hospital Israelita Albert Einstein, Brazil. 
     }%
 }

\begin{document}

\maketitle
\thispagestyle{empty}
\pagestyle{empty}

\begin{abstract}
Leader–follower interaction is an important paradigm in human–robot interaction (HRI). Yet, assigning roles in real-time remains challenging for resource-constrained mobile and assistive robots. While large language models (LLMs) have shown promise for natural communication, their size and latency limit on-device deployment. Small language models (SLMs) offer a potential alternative, but their effectiveness for role classification in HRI has not been systematically evaluated. In this paper, we present a benchmark of SLMs for leader–follower communication, introducing a novel dataset derived from a published database and augmented with synthetic samples to capture interaction-specific dynamics. We investigate two adaptation strategies: prompt engineering and fine-tuning, studied under zero-shot and one-shot interaction modes, compared with an untrained baseline. Experiments with Qwen2.5-0.5B reveal that zero-shot fine-tuning achieves robust classification performance (86.66\% accuracy) while maintaining low latency (22.2~ms per sample), significantly outperforming baseline and prompt-engineered approaches. However, results also indicate a performance degradation in one-shot modes, where increased context length challenges the model's architectural capacity. These findings demonstrate that fine-tuned SLMs provide an effective solution for direct role assignment, while highlighting critical trade-offs between dialogue complexity and classification reliability on the edge.
\end{abstract}

\section{INTRODUCTION}

As human-robot interaction (HRI) evolves in healthcare and assistive domains, systems are transitioning from executing simple commands to managing complex dyadic coordination~\cite{asadi2025redrawing, Adebayo2026dyadic, tschida2025exploring}. The primary goal of assistive robotics is to leverage user autonomy. However, effective assistance often requires the robot to intervene or take initiative when the user needs guidance or orientation. In scenarios such as indoor navigation in hospitals or clinics, the robot may need to guide the person to a location or be asked to accompany the person to help her when they reaches the desired location. Additionally, such systems can support remote homecare solutions, particularly in rural areas, for example for elderly coffee farmers in Brazil. In either case, the interaction is effectively modeled through the leader--follower paradigm~\cite{Takai2025leaderfollower}. In these contexts, role specialization enables the dyad to achieve superior task performance, provided that the robot can accurately determine when to lead, providing direction or proactive support, and when to follow, preserving the user's agency and intended pace~\cite{noormohammadi2025human, kwon2019influencing}. Once classified, this role serves as a high-level trigger for the robot's underlying control architecture, determining whether to execute a proactive guidance policy or a reactive, impedance-based following policy.

Establishing a robust communication channel for this mixed-initiative dialogue is essential to avoid coordination breakdowns and ensure therapeutic or supportive success~\cite{bonarini2020communication}. Natural language has emerged as a preferred interface for human-robot voice dialogue, as it reduces the unnatural feel of the interaction and increases user acceptance by allowing for fluid, conversational communication~\cite{zubiaga2024natural, cherakara-etal-2023-furchat}. While Large Language Models (LLMs) have shown significant promise in interpreting the linguistic ambiguity and context-dependent nature of such requests~\cite{wang2024lami, lim2024enhancing}, their deployment on mobile assistive platforms is frequently hindered by high latency, significant power consumption, and a reliance on stable internet connectivity~\cite{dhar2024empirical}.

\begin{figure}
    \centering
    \includegraphics[width=0.95\linewidth]{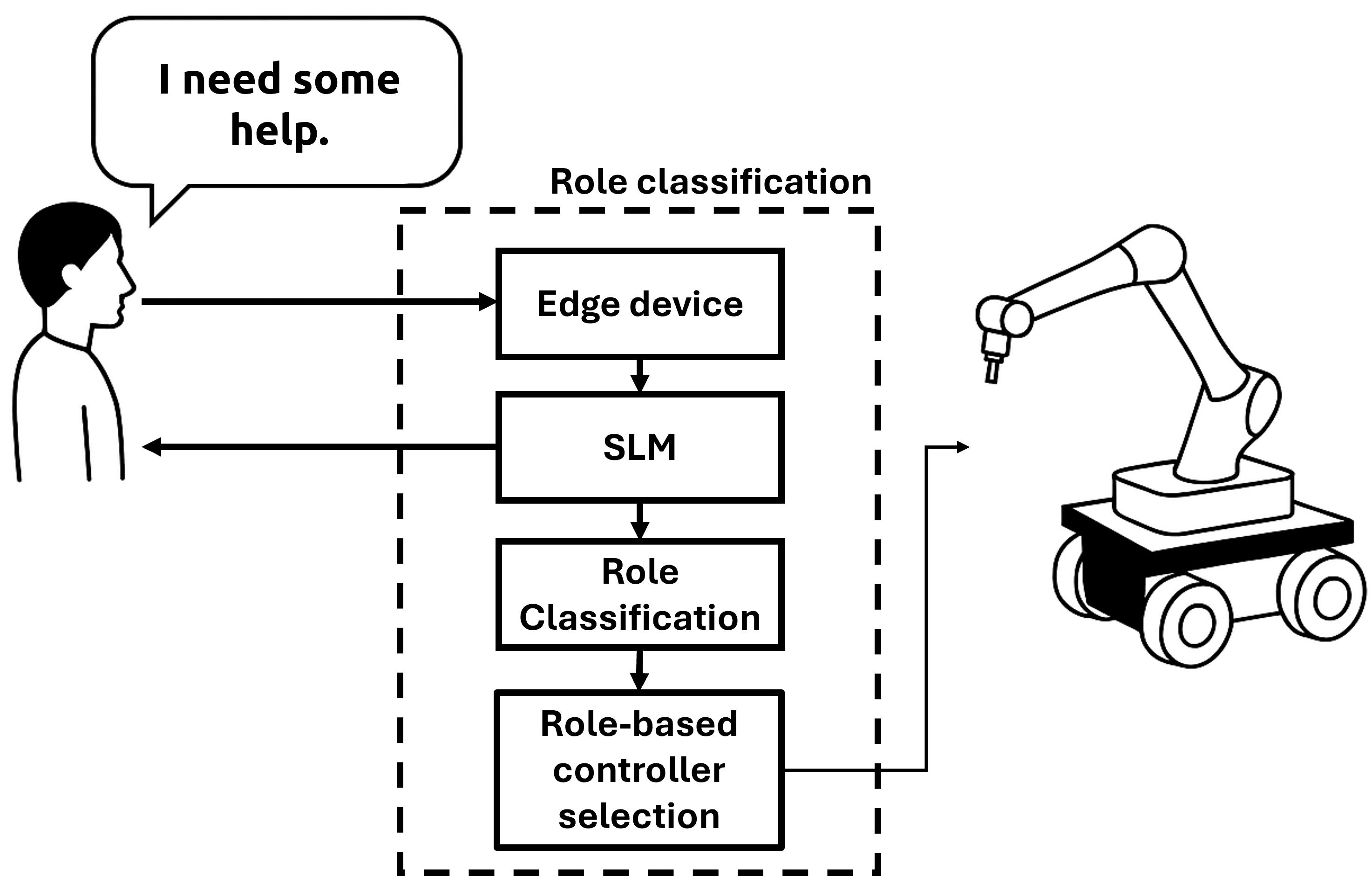}
    \caption{Edge-deployed leader–follower classification for assistive dyadic interaction. The system architecture selects the most appropriate low-level controller based on the social interation.}
    \label{fig:figure1}
\end{figure}

These constraints necessitate the adoption of Small Language Models (SLMs) optimized for edge computing. Unlike their larger counterparts, SLMs offer the benefits of being internet-independent and significantly faster, maintaining the real-time responsiveness required for safe and natural HRI without the prohibitive resource demands of massive architectures~\cite{kim_small_2025, yan2025we}.

Studies have assessed how SLMs perform on resource-constrained edge platforms, highlighting both their feasibility and limitations. Measurement and empirical analyses show that only small models (sub-4B parameters) run reliably on-device, with latency, memory, and thermal constraints severely limiting larger deployments; compression helps but often at the cost of quality~\cite{yan2025we,dhar2024empirical}. Systems-level research has further explored inference optimizations, such as speculative decoding and sparse mixture-of-experts, to mitigate latency and memory demands on embedded hardware~\cite{Xu2025EdgeLLM,10906629}. Application-driven comparisons on NVIDIA Jetson platforms also report that while Qwen2.5-3B is preferred when resources allow~\cite{Qwen225Report2024}, Qwen2.5-0.5B remains viable under tighter budgets, showing that carefully selected SLMs can deliver acceptable interactive performance onboard~\cite{JineteGomez2025TinyLLMsJetson, Qwen225Report2024}. Collectively, these studies confirm that SLMs can support real-time deployment on mobile robots, but also 
reinforces the persistent capacity–efficiency tensions of edge inference.

\begin{figure*}
    \centering
    \includegraphics[width=.9\linewidth]{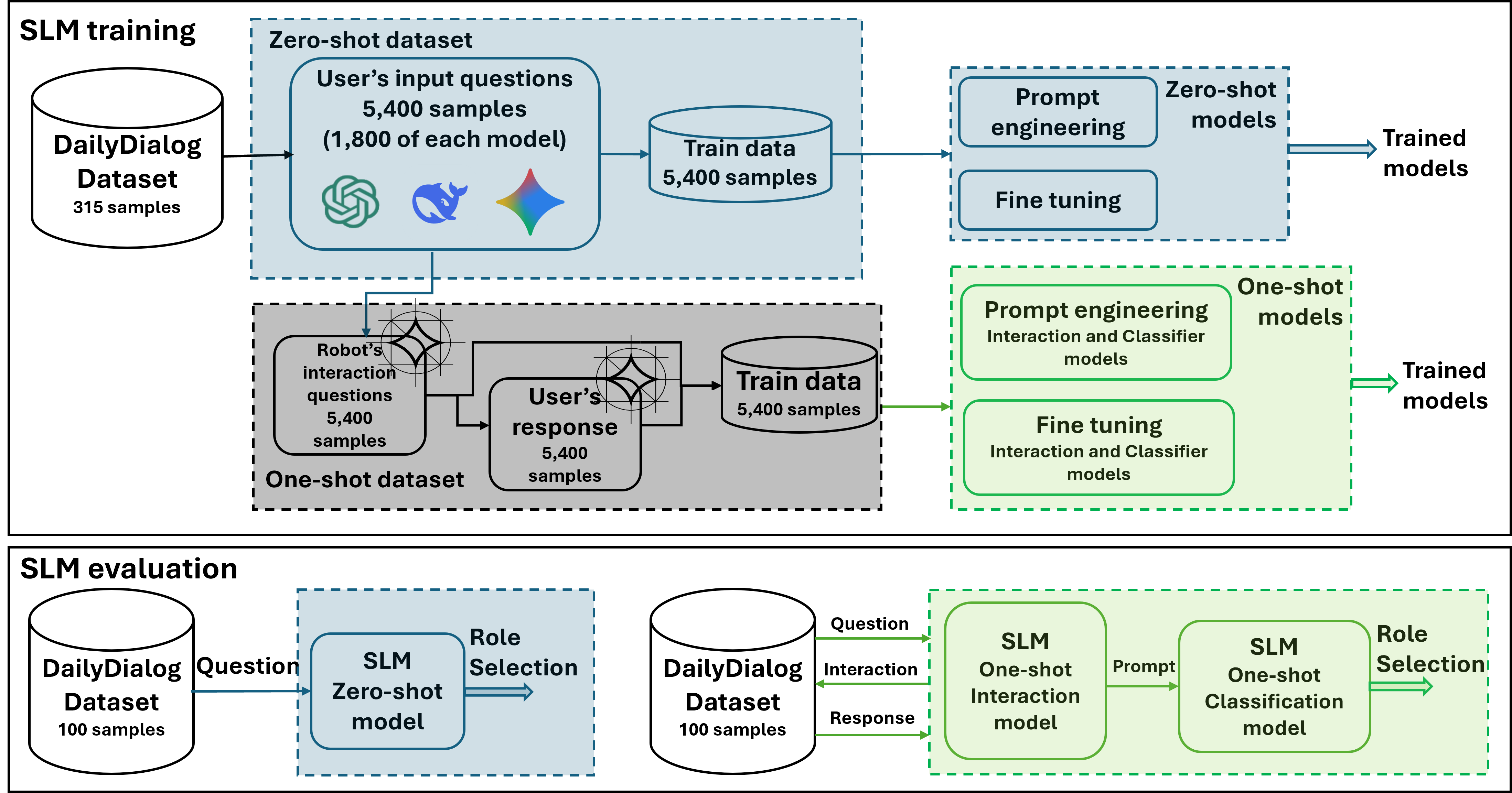}
    \caption{Overview of the dataset construction and model evaluation pipeline. The top section illustrates the creation of zero-shot and one-shot datasets, generated from DailyDialog samples and synthetic augmentation. The bottom section shows how these datasets were used to train and evaluate models under two adaptation strategies (prompt engineering and fine-tuning) for both zero-shot and one-shot interaction modes.}
    \label{fig:dataset_pipeline}
\end{figure*}

Despite these advances, the literature reveals important gaps in the assignment of leader–follower roles. Most edge-SLM studies emphasize general natural language processing (NLP) or systems metrics (throughput, memory, energy) rather than interaction-specific outcomes such as role inference fidelity, clarification behavior, or task success in closed-loop robot control~\cite{yan2025we,dhar2024empirical}. 
Moreover, to the best of our knowledge, there are no publicly available datasets specifically designed for leader–follower communication to support reproducible benchmarking of SLMs on embedded platforms, a gap also emphasized in the systematic review conducted by Corradini et al.~\cite{Corradini2025SLMSurvey}. 

Finally, while zero-shot and one-shot prompting strategies have been extensively studied in NLP~\cite{Sivarajkumar2023PromptZeroShotClinical,liu2023pre}, their implications for role assignment in leader–follower HRI remain underexplored. Prior work further demonstrates that task-specific prompting can outperform general-purpose approaches~\cite{ozsoy2024multilingual,mendoncca2023simple}, highlighting the need to compare prompting strategies with fine-tuning in this domain.

In this work, we address these gaps by conducting a systematic evaluation of an SLM within the leader–follower HRI context, as shown in Fig.~\ref{fig:figure1}. Our contributions are twofold: (1) we introduce and make publicly available novel datasets tailored to leader–follower communication in HRI, enabling systematic training and evaluation of models; and (2) we perform a comparative study of prompt engineering and fine-tuning strategies for SLM adaptation on edge devices, explicitly contrasting zero-shot and one-shot prompting with fine-tuned approaches to analyse their respective strengths and limitations. 


\section{METHODS}

This section details the methodological framework adopted in our study. We built our method on processes of synthetic data generation to augment our dataset~\cite{nadas_synthetic_2025, ren_few-shot_2025, josifoski_exploiting_2023} and experimental comparisons between prompt engineering and fine-tuning~\cite{zhang_comparison_2024}. In the remainder of this section, we first describe the construction of a leader–follower dataset derived from existing dialogue resources and augmented with synthetic samples. Next, we present the rationale for selecting Qwen2.5-0.5B as the base SLM and outline the interaction modes considered (zero-shot and one-shot). We then introduce two adaptation strategies: prompt engineering, where we design zero-shot and one-shot prompt architectures, and fine-tuning, where the model is further specialized for binary role classification.

\subsection{Dataset} \label{sec:dataset} 
As highlighted in the introduction, we conducted extensive searches but found no publicly available datasets specifically tailored to leader–follower communication in HRI\footnote{Searches in February 2026.}. To address this gap, we created our own dataset\footnote{Datasets and prompts are available in our \href{https://github.com/bme-research/biorob2026_slm_leader_follower}{GitHub repository}.} by leveraging samples from the dialogue and intent classification dataset DailyDialog~\cite{li2017dailydialog}. We extracted 415 questions that aligned with leader–follower dynamics. The selection was based on identifying questions that required guidance or orientation, and each label was assigned according to the nature of the dialogue in the original context. If one of the participants in the dialogue requested guidance or directions, the label assigned was \texttt{LEADER}. Conversely, if a participant initiated a task alone or suggested being accompanied, the label assigned was \texttt{FOLLOWER}. 

To augment the dataset, we utilized 315 selected questions (148 \texttt{FOLLOWER} and 167 \texttt{LEADER}), while reserving 100 original human-sourced questions (50 \texttt{FOLLOWER} and 50 \texttt{LEADER}) for model evaluation. The data augmentation process was conducted using three different LLMs (DeepSeek, Gemini, and GPT-4) following methods established in recent research on synthetic data generation with LLMs~\cite{long2024llms, nadas_synthetic_2025, ren_few-shot_2025, josifoski_exploiting_2023}. Each model was instructed to generate approximately six paraphrases per question, preserving the original samples' meaning and structure~\cite{ding2024data}. This resulted in a total of 5,400 leader-follower-oriented questions, 1,800 per model. This dataset size is consistent with related works that employed a similar number of samples~\cite{griesshaber2020fine, alex2021raft}.

Two dataset configurations were derived to align with the interaction modes tested in this study. The first corresponds to a \textit{zero-shot} setup and contains leader–follower questions with their associated role labels. The second corresponds to a \textit{one-shot} setup and simulates more complex interactions. It includes the generated questions, an additional clarifying question about the user’s intentions, and a response generated by an LLM simulating a human-like but intentionally ambiguous reply. This “scarecrow” validation methodology~\cite{williams2024scarecrows}, replaces the need for a human interlocutor during early-stage validation by acting as a temporary black-box LLM component within the system pipeline. This design enables scalable and reproducible experimentation. Moreover, it mitigates ethical risks associated with exposing real participants to potentially misaligned or unvalidated behaviors, functioning as an intermediate validation layer before transitioning to human-subject studies ~\cite{wen2024gpt}. 

To evaluate the quality and semantic integrity of the augmented datasets, a semantic fidelity analysis was conducted by comparing the synthetic samples against the original human-annotated data. We employed the Sentence-BERT (SBERT) framework to generate high-dimensional embeddings for both real and synthetic instances~\cite{chang2024survey}. For each synthetic sentence within a specific class, we calculated the maximum cosine similarity relative to all real samples in the same category. This approach ensures a balanced assessment of how well the synthetic generation preserves the core semantic meaning of the original classes, with results interpreted based on the distribution of maximum similarity scores. 
The mean maximum cosine similarity for the GPT-4 augmented dataset was $0.831 \pm 0.086$ with respect to the \textit{Follower Rule} and $0.846 \pm 0.090$ with respect to the \textit{Leader Rule}. For the Gemini-augmented dataset, the corresponding values were $0.786 \pm 0.107$ and $0.788 \pm 0.120$, respectively. The DeepSeek-augmented dataset achieved $0.826 \pm 0.097$ under the Follower Rule and $0.828 \pm 0.113$ under the Leader Rule.

Finally, each interaction in both datasets was labeled with the correct role assignment, establishing a ground truth for evaluation. The overall process of dataset construction and organization is summarized in Fig.~\ref{fig:dataset_pipeline}. To our knowledge, this represents the first dataset specifically tailored to leader–follower HRI communication, enabling systematic benchmarking of SLM adaptation strategies.

\begin{table*}[t] \caption{Performance comparison of evaluated models across zero-shot and one-shot interaction modes. Metrics include classification accuracy, precision, recall, and F1-score, alongside efficiency measures (tokens per second and latency).}
\renewcommand*{\arraystretch}{1.5} 
\centering
\resizebox{\textwidth}{!}{
\begin{tabular}{llccccccc}
\hline
\textbf{Interaction}                & \textbf{Method}      & \textbf{Accuracy} [\%]   & \textbf{Precision} [\%] & \textbf{Recall} [\%] & \textbf{F1-score} [\%] & \textbf{Tokens/s}         & \textbf{Latency} [ms] \\ \hline
\multirow{3}{*}{\textbf{Zero-shot}} & \textbf{Baseline}    & 55.00$\pm$10.98          & 75.17$\pm$21.61         & 16.39$\pm$6.53       & 25.99$\pm$8.23         & 281.9$\pm$105.8           & 33.8$\pm$3.7                       \\
                                    & \textbf{Prompt Eng.} & 53.87$\pm$10.38          & 56.80$\pm$19.20         & 31.03$\pm$12.58      & 38.80$\pm$12.66        & 85.9$\pm$27.0             & 111.9$\pm$40.0                         \\
                                    & \textbf{Fine-tuning} & 86.66$\pm$6.77           & 84.81$\pm$10.07         & 90.05$\pm$7.98       & 86.72$\pm$6.53         & 432.1$\pm$148.2           & 22.2$\pm$4.2             \\ \hline
\multirow{3}{*}{\textbf{One-shot}}  & \textbf{Baseline}    & 48.00$\pm$9.09           & 40.83$\pm$32.16         & 9.27$\pm$7.99        & 14.18$\pm$11.39        & 988.63$\pm$116.7          & 41.0$\pm$2.6                        \\
                                    & \textbf{Prompt Eng.} & 45.07$\pm$8.17           & 42.48$\pm$26.49         & 17.41$\pm$10.91      & 23.12$\pm$13.42        & 305.6$\pm$90.7            & 213.8$\pm$309.1                  \\
                                    & \textbf{Fine-tuning} & 51.65$\pm$13.40          & 43.69$\pm$45.02         & 8.90$\pm$13.29       & 13.07$\pm$16.55        & 1851.69$\pm$380.09        & 22.2$\pm$3.5             \\ \hline
\end{tabular}} \label{tab:main_results}
\end{table*}

\subsection{Language Model Evaluation}
Based on the constraints outlined in the introduction, we required an SLM that balanced computational efficiency with sufficient accuracy for leader–follower role assignment. The literature shows that compact models from the Qwen2.5 family achieve strong parameter efficiency, with the 0.5B variant delivering competitive comprehension and instruction-following performance despite its reduced scale~\cite{yan2025we}. Benchmark results reported in the Qwen2.5 technical report further demonstrate that this variant outperforms other models of comparable size across standard evaluations while maintaining low latency, reinforcing its suitability for deployment on resource-constrained robotic platforms~\cite{Qwen225Report2024}. These findings support our choice of Qwen2.5-0.5B as the base model. Its balance of computational efficiency and task performance makes it well-suited for embedded assistive robotics, where real-time responsiveness and resource efficiency are critical.

We evaluated this model under two \textit{interaction modes}: zero-shot and one-shot. In zero-shot mode, the SLM must classify the leader–follower role using only the user’s initial input, with no opportunity for clarification. In the one-shot mode, the SLM is permitted to ask a single clarifying question, after which it combines the original input and the user’s response to make its prediction. The bottom half of Fig.~\ref{fig:dataset_pipeline} illustrates how these datasets are integrated into the evaluation pipeline, where zero-shot and one-shot models are trained and assessed using prompt engineering and fine-tuning strategies. Together, these modes provide a framework for systematically evaluating the impact of clarification on role assignment. The performance of all proposed strategies was evaluated using Monte Carlo Cross-Validation (MCCV), with 30 independent iterations to ensure statistical de-biasing. We next describe the baseline and the two adaptation strategies applied to this model: prompt engineering and fine-tuning. As a reference condition, the Qwen2.5-0.5B model, without any task-specific adaptation, was used as baseline.

\subsubsection{Baseline Model}
We employed Qwen2.5-0.5B model with its original weights as the baseline, without applying few-shot examples or any task-specific adaptation. The model received only the task instruction and generated responses based on its pre-trained parameters.

\subsubsection{Prompt Engineering}
We designed prompt architectures for both the zero-shot and one-shot setups. The zero-shot architecture employs a single system prompt that encapsulates the agent's personality, behavior, and task-specific instructions, drawing on prompt engineering methods such as few-shot prompting and meta-prompting~\cite{chang2024survey,gao2021making,liu2023pre}. 

In contrast, the one-shot architecture uses two system prompts, each tailored to a specific functional component of the agent. The first prompt is designed to elicit a clarifying question from the model. In contrast, the second integrates both the user’s initial input and their clarification response to determine the final role assignment. This decomposition was motivated by the limitations of small-scale models, where breaking instructions into smaller, focused prompts can improve consistency and overall performance~\cite{gao2021making,liu2023pre}.

\subsubsection{Fine-tuning}
While prompt engineering can adapt an SLM to specific tasks, its effectiveness is often constrained by the model’s pretraining distribution, especially in small-scale models~\cite{gao2025llm,chung2024instruction}. To address this limitation, we fine-tuned the Qwen2.5-0.5B model on our generated dataset to specialize it for binary leader–follower role classification. Fine-tuning has been widely adopted to improve task adherence when zero-shot or few-shot prompting alone proves insufficient. The fine-tuning process was conducted on an NVIDIA server equipped with L40S 48GB GPUs.

For this step, we used the Autotrain framework for text classification~\cite{parthasarathy2024ultimate}, which we chose for its practicality and efficiency in hyperparameter exploration. In this framework, the base model is adapted for the classification task by appending a lightweight linear head to the pretrained architecture. This head learns to map the model’s latent representations into discrete labels, bridging the gap between general-purpose language understanding and the specific requirements of leader–follower classification. The parameters used on the framework are: 10 training epochs, a learning rate warmup of 50 steps, weight decay of 0.01, and a batch size of 16 for both training and evaluation. The model was optimized using mixed-precision (FP16) training, with the best model selected based on accuracy and a stratified 20\% validation split.

In the zero-shot condition, a single fine-tuned model is sufficient to directly classify the role based on the user’s initial request. In the one-shot condition, however, we trained two complementary models: one fine-tuned to generate a clarifying question when the initial input is ambiguous, and a second fine-tuned to perform the final leader–follower classification using both the original input and the user’s follow-up answer. This separation ensures that the clarification stage is optimized for interaction quality, while the classification stage remains optimized for decision accuracy.


\section{EXPERIMENTAL EVALUATION}

This section presents the evaluation outcomes of the proposed leader–follower classification framework. We first report the overall performance of different adaptation strategies (baseline, prompt engineering, and fine-tuning) across zero-shot and one-shot interaction modes. Each proposed SLM model was evaluated on the test set of 100 phrases, reporting key performance metrics such as overall accuracy, per-class precision, recall, and F1-score, together with system efficiency in terms of throughput and latency. To ensure statistical robustness, this evaluation was repeated thirty times, allowing us to run statistical analysis on the results. Experiments were conducted under both zero-shot and one-shot conditions. Based on these results, we identify the best-performing model and, in a subsequent subsection, perform a focused analysis on the impact of sentence length on classification performance.

\subsection{Classification Metrics}

\begin{figure}
    \centering
    \includegraphics[width=\linewidth]{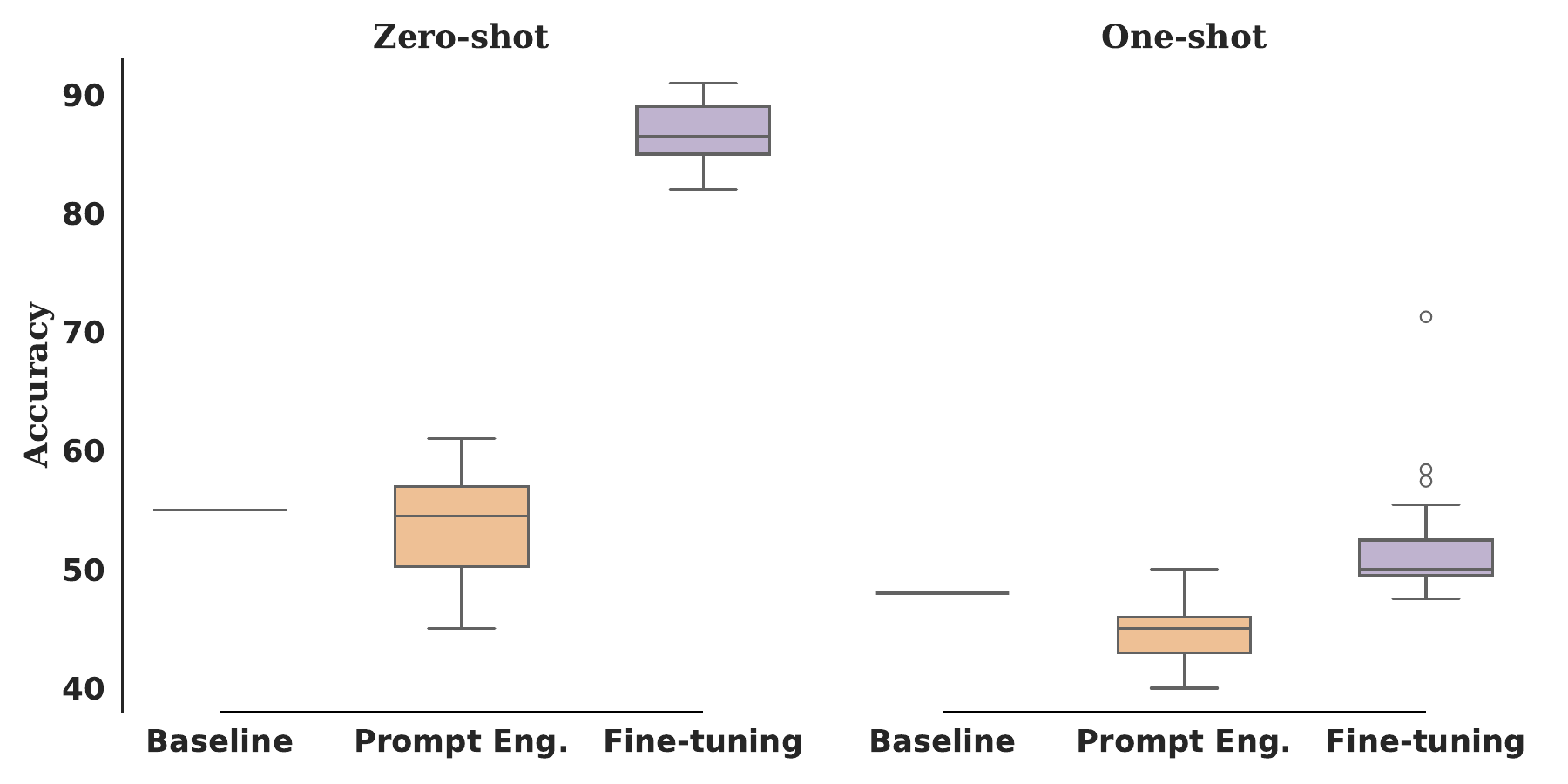}
    \caption{Accuracy dispersion across 30 runs for each interaction mode and adaptation strategy. Fine-tuned models consistently achieve higher median accuracy than baseline and prompt-engineered approaches, though the one-shot fine-tuned condition shows slightly greater variability.}
    \label{fig:accuracy_boxplot}
\end{figure}

Table~\ref{tab:main_results} summarizes the results across methods and interaction modes. For the zero-shot interaction, fine-tuning significantly outperformed both the baseline and prompt engineering strategies. In this setup, fine-tuning achieved an accuracy of 86.66\%, compared to 53.87\% for prompt engineering and 55.00\% for the baseline (z-test, $p<0.001$). A different pattern was observed in the one-shot setup, where performance across all methods converged near chance level; fine-tuning reached only 51.65\% accuracy, statistically comparable to the baseline (48.00\%) and prompt engineering (45.07\%) (z-test, $p>0.001$). Precision and recall metrics reveal that while zero-shot fine-tuning achieves a robust balance, the zero-shot baseline and prompt-engineering methods are significantly skewed toward higher precision at the expense of extremely low recall (e.g., 16.39\% for the baseline). This suggests that without specialization, the models are overly conservative, failing to detect the majority of leader–follower transitions. In the one-shot configuration, however, fine-tuning failed to maintain this performance, with recall dropping to a marginal 8.90\%. This shift indicates that, while fine-tuned SLMs are highly effective for direct classification, the increased semantic complexity and context length of one-shot interactions—incorporating both the clarifying question and the simulated response—may exceed the 0.5B model's limited parameter capacity, leading to a breakdown in classification.

\begin{figure*}[t] 
    \centering
    \begin{subfigure}[b]{0.48\linewidth}
        \centering
        \includegraphics[width=\linewidth]{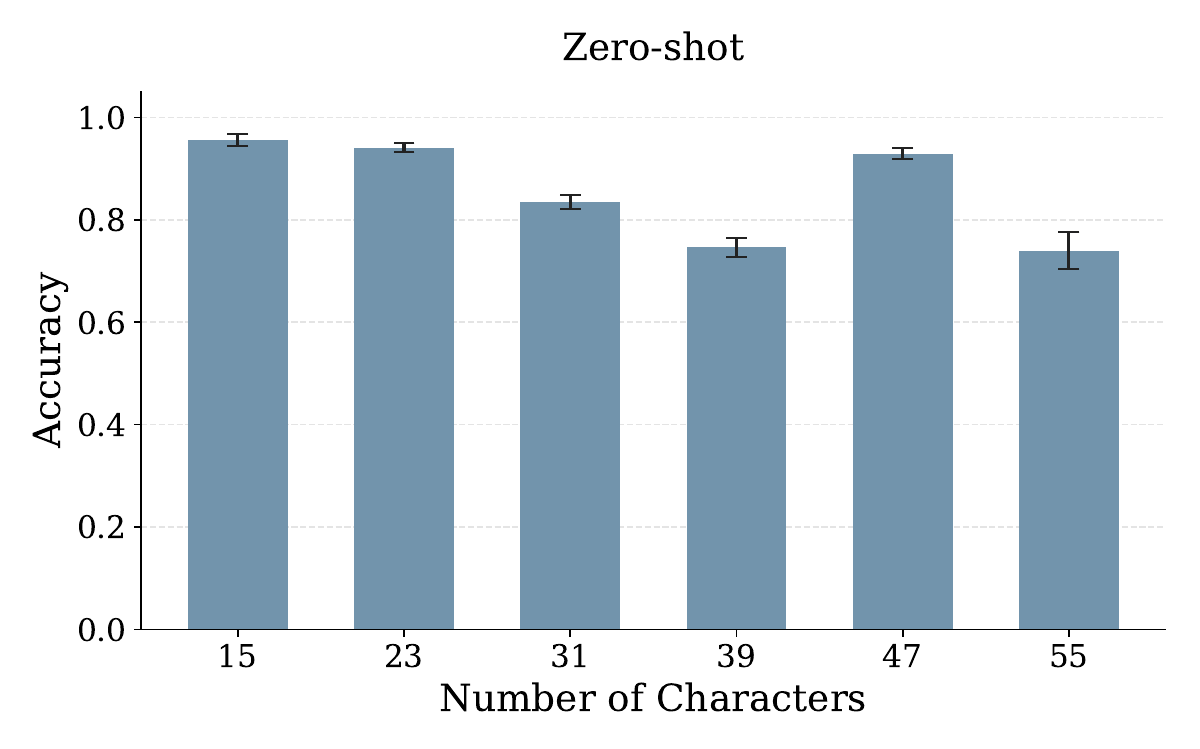}
        \caption{}
        \label{fig:accuracy-vs-sentence-length_zero}
    \end{subfigure}
    \hfill
    \begin{subfigure}[b]{0.48\linewidth}
        \centering
        \includegraphics[width=\linewidth]{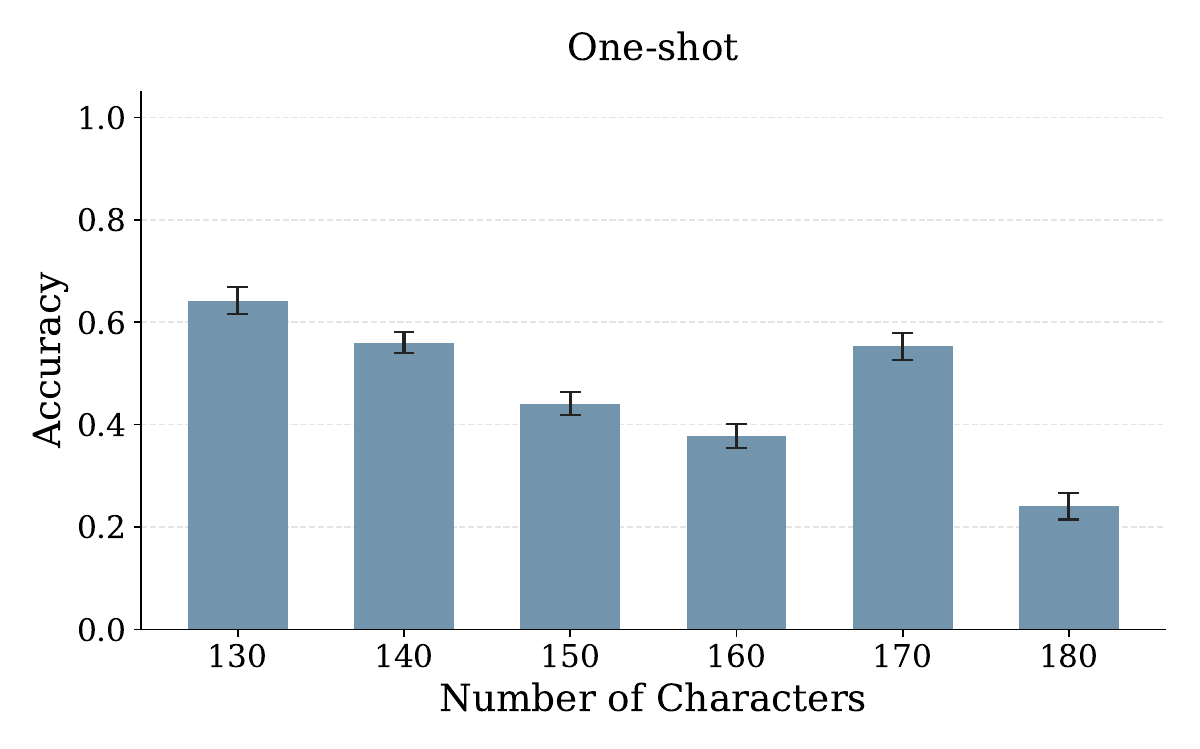}
        \caption{}
        \label{fig:accuracy-vs-sentence-length_one}
    \end{subfigure}
    \caption{Accuracy as a function of sentence length for the fine-tuned models under (a) zero-shot and (b) one-shot settings. Error bars indicate one standard deviation. For the zero-shot setting, sentence length is measured using only the user's question. In the one-shot setting, sentence length is computed from the concatenation of the user's question, the provided example answer, and the final response, resulting in a larger number of characters overall.
    }
    \label{fig:accuracy_vs_length_combined}
\end{figure*}

When comparing zero-shot and one-shot modes, the fine-tuned model's performance degraded significantly, with accuracy dropping from 86.66\% to 51.65\% ($p<0.001$). Similarly, the baseline accuracy declined from 55.00\% to 48.00\%, while prompt engineering results decreased from 53.87\% to 45.07\%. The dispersion observed in Fig.~\ref{fig:accuracy_boxplot} suggests that the one-shot interaction mode introduces a level of linguistic complexity that the 0.5B parameter model struggles to process reliably. This is most evident in the fine-tuned one-shot metrics, where precision fell to 43.69\% and recall collapsed to a marginal 8.90\%. These results indicate that while fine-tuned SLMs are highly effective for direct, single-turn role classification, the increased context length—incorporating both a clarifying question and a synthetic response—likely exceeds the model's capacity to maintain semantic fidelity. In leader–follower HRI, this is an important finding: it suggests that for small models on the edge, a more complex, multi-turn dialogue may actually increase the risk of coordination breakdowns by introducing noise that the model cannot successfully filter to determine initiative. Further research into specialized context-pruning pipelines is necessary to enable the linguistic flexibility of multi-turn dialogues without compromising the system's initiative-detection reliability.

Finally, efficiency analysis revealed clear trade-offs between adaptation strategies. Fine-tuning consistently achieved the lowest latency (around 22~ms per sample) and the highest throughput (432--1851 tokens/s), significantly outperforming both baseline and prompt engineering. Prompt engineering incurred the highest latency, particularly in the one-shot mode (213.8~ms), due to the added prompt complexity and the need to maintain longer context windows within the 0.5B model's inference cycle. Interestingly, while the one-shot configuration produced more tokens on average—due to the interaction including both a clarifying question and a synthetic response—the measured latency for the fine-tuned model remained identical to the zero-shot case (22.2~ms). Due to high throughput, fine-tuned one-shot models incur negligible latency overhead compared to zero-shot modes. However, the trade-off is not negligible for prompt-engineered approaches, where latency increased nearly twofold in one-shot mode.

\subsection{Sentence Length Analysis}

Given recent reports that longer inputs can exacerbate language model hallucinations~\cite{chakraborty2025hallucination}, we evaluated whether sentence length affects the reliability of leader–follower classification. Specifically, we tested whether increased context in longer inputs improved decision quality or increased the model's susceptibility to misclassification. To analyze this, we performed data binning on the test set with a step size of 10 characters, calculating the mean accuracy and standard deviation across thirty independent runs for each bin. The x-axis of Fig.~\ref{fig:accuracy_vs_length_combined} indicates the character length bins, while the y-axis represents the classification accuracy.

The results reveal distinct patterns for the two interaction modes. In the zero-shot configuration (Fig.~\ref{fig:accuracy-vs-sentence-length_zero}), accuracy remained robust across most bins, maintaining levels above 80\% for shorter sentences and showing relative stability even as length increased, despite a localized dip in the 39-character bin. This suggests that the 0.5B model effectively captures role intent in single-turn interactions regardless of moderate phrasing complexity. By contrast, the one-shot configuration (Fig.~\ref{fig:accuracy-vs-sentence-length_one}) exhibited a significant and progressive decline in performance. Accuracy dropped from approximately 65\% in the shortest bin to nearly 25\% for the longest sentences.

Taken together, these findings indicate that while zero-shot classification is relatively resilient to sentence length, one-shot interactions struggle with the combination of longer inputs and multi-turn context leads to a breakdown in semantic fidelity. For extremely small models on the edge, the increased noise of a verbose, multi-turn dialogue appears to overwhelm the model's parameters, underscoring the importance of concise interaction design when using sub-1B parameter models for initiative-switching.

%


\section{CONCLUSION}


This work presented an evaluation of SLMs for leader–follower role classification in HRI. We constructed a novel dataset tailored to this task, introduced two interaction modes (zero-shot and one-shot), and compared prompt engineering and fine-tuning strategies on the Qwen2.5-0.5B model. Across thirty independent trials, zero-shot fine-tuning consistently outperformed prompt engineering and baseline methods in both accuracy and efficiency. However, one-shot interactions exhibited a significant performance degradation, highlighting critical architectural limits when processing multi-turn context on extremely small models. Together, these results reveal trade-offs among efficiency, precision, and dialogue complexity when deploying SLMs on edge-constrained robotic platforms.

Beyond empirical performance, our study makes two main contributions to the field. First, we provide the first publicly available dataset designed specifically for leader–follower HRI communication, enabling reproducible benchmarking of adaptation strategies. Second, we demonstrate that fine-tuning is not only superior in accuracy but also more efficient in terms of latency and throughput, directly addressing the practical constraints of embedded robotic platforms. In this context, our sentence length analysis revealed that while one-shot configurations are less sensitive to input complexity and length than zero-shot models, they require greater parameter capacity or specialized context management to maintain classification fidelity.

Despite the observed poor performance in 0.5B models, one-shot interactions retain significant potential as they closely mimic the inherent complexity of real-life dyadic coordination. To transition from direct role assignment to these more natural dialogue paradigms, future research must develop specialized evaluation frameworks for the edge. Such frameworks are needed to assess the semantic quality of the SLM's clarifying questions and the contextual veracity of the scarecrow simulation responses. This necessitates a shift toward either rigorous mathematical evaluations of semantic fidelity or the inclusion of subjective user judgment through human-in-the-loop studies.

There remain, however, important directions for future work. Our dataset, while novel, is synthetic and limited in scale; extending it with larger, multimodal, and human-annotated samples will strengthen the ecological validity of the findings. Similarly, exploring alternative SLM architectures, compression strategies, and multilingual settings may further improve efficiency and inclusion. Finally, evaluating these models in real-world robotic experiments, where communication is embedded in dynamic tasks and environments, will be critical to validate their robustness and user acceptance. Future research will also address the underexplored potential of one-shot interactions by conducting deeper evaluations of the SLM-generated answers and the simulation (scarecrow) module's performance, using both automated metrics and human-in-the-loop studies to refine multi-turn dialogue reliability.

\section*{ACKNOWLEDGMENT}
\small
This study was financed, in part, by the São Paulo Research Foundation (FAPESP), Brasil, process Number 2025/09390-3, and by the Hospital Israelita Albert Einstein's Seed Money 2024 call. We also acknowledge Minas Gerais State Research Support Foundation (FAPEMIG) - APQ-00996-24; \textit{CAPES}: Coordination for the Improvement of Higher Education Personnel -- Brazil (CAPES), via institutional funding (Code 001); UFLA -- Universidade Federal de Lavras, funding N 014/2026 -- INOV\@AÇÃO  process Number 23090.000597/2026-73;  2024 Cultural Technology Research and Development Project by the Ministry of Culture, Sports and Tourism and the Korea Creative Content Agency(Project Name: Development of AX Authoring Technology and Immersive Content Based on MR Spatial Computing, and Global Specialist Training, Project Number: RS-2024-00399186, Contribution Rate: 100\%); and \textit{CNPq}: National Council for Scientific and Technological Development (CNPq), for partial financial support. Finally, this work was also supported by the Petr\'{o}leo Brasileiro S/A - Petrobras,
using resources from the R\&D clause of the ANP, in partnership with the
Universidade de S\~{a}o Paulo (USP) and the Funda\c{c}\~{a}o de Apoio \`{a} F\'{\i}sica e
\`{a} Qu\'{\i}mica (FAFQ), under Cooperation Agreement No. 2023/00016-6 and
2023/00013-7.


\bibliographystyle{IEEEtran}
\bibliography{biblio}

\end{document}